\documentclass[prl,aps,%
,twocolumn%
,secnumarabic
,amssymb]{revtex4}
\usepackage{amsmath}%
\usepackage{bm}%
\usepackage{multirow}
\usepackage{ifpdf}%
\usepackage{epsf}%
\ifpdf
\usepackage[pdftex]{graphicx}
\else
\usepackage{graphicx}
\fi

\usepackage{color}

\ifpdf
\DeclareGraphicsExtensions{.pdf, .jpg}
\else
\DeclareGraphicsExtensions{.eps, .jpg}
\fi

\begin{document}
\title{Energy Dissipation of Fast Electrons in Polymethylmetacrylate (PMMA):\\Towards a Universal Curve for Electron Beam Attenuation in Solids \\for Energies between $\sim$0~eV and 100~keV}
\author{ Wolfgang S.M. Werner\footnote{werner@iap.tuwien.ac.at, fax:+43-1-58801-13499}}%
\author{Florian Simperl}%
\author{Felix Bl{\"o}dorn}%
\author{Julian Brunner}%
\author{Johannes Kero}%
\affiliation{Institut f\"ur Angewandte Physik, Technische Universit\"at  Wien, Wiedner Hauptstra{\ss}e 8-10/E134, A-1040 Vienna, Austria}
\author{Alessandra Bellissimo}
 \affiliation{Institut f\"ur Photonik, Technische Universit\"at  Wien, Gu{\ss}hausstra{\ss}e 27-29/E387, A-1040 Vienna, Austria}
\author{Olga Ridzel}
 \affiliation{Theiss Research, 7411 Eads Ave., La Jolla, CA 92037-5037, USA.}

\begin{abstract} 
Spectroscopy of correlated electron pairs was employed to investigate the energy dissipation process as well as the 
transport and the emission of low energy electrons on a polymethylmetracylate (PMMA) surface,
providing secondary electron (SE) spectra causally related to the energy loss of the primary electron. 
Two  groups of electrons are identified in the cascade of slow electrons, corresponding to different stages in the energy dissipation process. 
For both groups, the characteristic lengths for attenuation due to collective excitations and momentum relaxation
are quantified and are found to be distinctly different: $\lambda_1=(12.0\pm2)\rm{\AA}$ and  $\lambda_2=(61.5\pm11)\rm{\AA}$. 
The results strongly contradict the commonly employed model  of exponential attenuation 
with the electron inelastic mean free path (IMFP) as characteristic length, but essentially agree with a theory  used for  decades in astrophysics 
and neutron transport, albeit with characteristic lengths expressed in units of {\rm{\AA}}ngstr{\o}ms rather than lightyears.
\end{abstract}
\maketitle 

\noindent 
Electrons with vacuum energies in the range of $\sim$0-20~eV are playing an increasingly important role in modern science and technology. While low energy electrons (LEEs) 
have been utilised for a century in electron microscopy  \cite{goldstein}, modern applications of  nanotechnology require an improved understanding of the energy dissipation of LEEs in solid  surfaces. This  concerns the effective interaction volume in electron beam lithography caused by electron diffusion (proximity effect)    \cite{kozawa2010,kozawa2021,proximity,euvossiander} as well as focussed electron beam 
deposition   \cite{febid}, spacecraft surface charging  \cite{garret}, electron cloud formation in charged particle storage rings  \cite{ciminoprl93,ohmiprl75} and plasma-wall interaction in fusion research  \cite{schourev}. LEEs are also the essential agents for DNA-strand breaks as a result of irradiation of biological tissue with ionising radiation  \cite{sanchescience}.  The transport of LEEs near solid surfaces is particularly important for the emerging fields of plasmonics  \cite{brongersmaplasmonics}  and photonics, e.g. to quantify photoelectron delay times due to collective excitations in solids for attosecond 
physics on solid surfaces  \cite{cavalierinature,schultzeatto,signorellprl,zewaildynamics}  or optical-field induced correlated electron emission  \cite{hommelhofnature}.

For medium energies  ($\sim$100~eV--100~keV), the electron--solid interaction   relevant to electron spectroscopy for surface analysis is nowadays quantitatively  
understood  \cite{wernerfrontiers}.
A case in point is the attenuation of electron beams penetrating a surface. Jab{\l}onski and Powell  \cite{jabpoweal} have recently reviewed the  development  of a method to reliably quantify electron attenuation, which took several decades. The commonly accepted model is an exponential 
attenuation law, with the  electron inelastic mean free path (IMFP, $\lambda_i$) as characteristic length, slightly modified  to account for the influence of  elastic electron scattering  in 
accordance with the employed experimental conditions. Measurements of the IMFP using electron reflection experiments agree quantitatively with theoretical 
results based on optical data and linear response theory  \cite{powjabepes,werjpcrd,werimfp2022}. 

At low energies ($<100$~eV), however, it is still not possible  to  satisfactorily describe essential observables  upon impact of a primary electron, such as the spectrum 
of emitted secondary electrons  (SEs) or the SE-yield, since additional physical  phenomena come into play that make the parameters of theoretical models less reliable,
while  experiments with LEEs are generally more difficult  \cite{alediss}. Refinement of any model  is complicated by the lack of benchmark experiments
specifically designed to obtain information on individual physical parameters or processes.

Concerning the length scale over which low energy electrons are attenuated, many authors  adopt the same approach as for medium energies, i.e. exponential attenuation with 
the IMFP as characteristic length.
The present work challenges  this approach for low energies.
We investigate the energy dissipation of fast electrons  in polymethylmetacrylate (PMMA), a 
photoresist commonly used in electron beam lithography \cite{kozawa2010,kozawa2021}, and study the transport and emission of  LEEs liberated upon impact of the primary.
Correlated electron pairs of primary (medium-energy) electrons striking a surface and secondary (low-energy) electrons 
emitted as a result are measured in coincidence, yielding secondary electron spectra causally related to a given energy loss of the primary 
after a certain number of inelastic collisions. The quantitative  model  for  medium-energy electron transport   
 \cite{powjabepes,wernerfrontiers,werjpcrd} is then invoked to  calculate the average depth at which a given number of 
energy losses of the  primary electrons take place. Since each energy loss creates 
a single secondary electron  \cite{wercoisec,weralcoinc,alebruce,werhopg}, comparison of the intensity of energy losses of the primary, i.e. the number of secondary 
electrons  {\em created at a certain depth} with the number {\em emitted into vacuum}, then provides the length scale over which  low energy secondary electrons are attenuated.

The results strongly contradict the commonly used exponential attenuation law.
 This observation is not unexpected, given the dynamic interplay between energy fluctuations arising from collective excitations (governed by the IMFP) and momentum relaxation attributed to elastic scattering by the Coulomb potential of the ionic cores (described by the transport mean free path 
($\lambda_{tr}$,TrMFP) \cite{werqsasia}). This relationship changes dramatically at energies below 100 eV.  
A universal attenuation law adequately accounting for these phenomena   developed earlier 
in astrophysics and neutron transport theory  \cite{case,chandrasekhar,davison}  describes our results satisfactorily. The present findings may thus help to resolve the ongoing 
controversy (see e.g. Refs.~ \cite{chantlerprl,devera2019,geelenprl}) regarding low energy electron beam attenuation in solids.

The  chain of processes we identify in the energy dissipation mechanism is expected to be more generally 
encountered, e.g.,  in biological matter exposed to ionising radiation \cite{sanchescience},
since the relevant electron--solid interaction characteristics and electronic structure   is similar for a large class of materials
\cite{werqsasia}.
\begin{figure}[t]
{\includegraphics[width=1.\columnwidth,trim={0.2cm 4.0cm 15.cm .cm},clip]{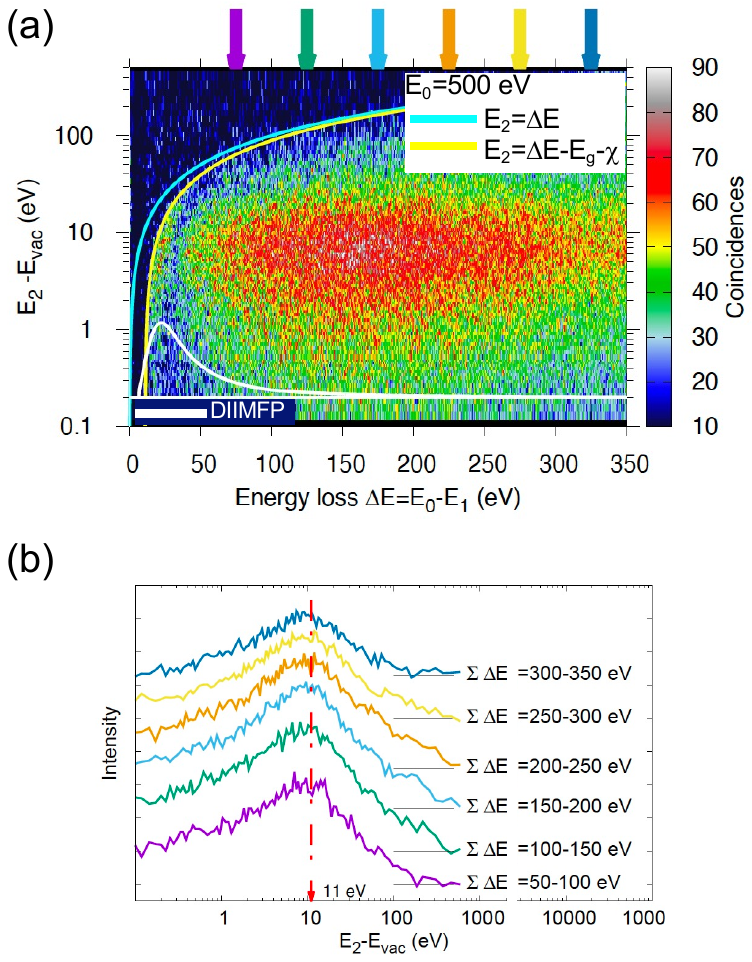}}
\caption{%
(a) Double differential secondary electron-electron energy loss coincidence spectra (SE2ELCS) for  $E_0=500$~eV electrons striking a PMMA surface.
The cyan curve indicates energies $E_2$ equal to the energy loss $\Delta E=E_0-E_1$ of the fast electron;
Yellow curve: maximum vacuum energy for an emitted electron  created by an energy loss $\Delta E$ of the primary after overcoming the surface barrier $U=E_g+\chi$.
White curve: differential inverse inelastic mean free path (DIIMFP).  
(b) SE spectra obtained by integrating the data in (a) over   indicated ranges of $\Delta E$  (see the arrows in (a)).
}
\label{f500}
\end{figure}
%

\begin{figure}[htb]
{\includegraphics[width=0.75\columnwidth,trim={0.2cm 11.0cm 10.cm .cm},clip]{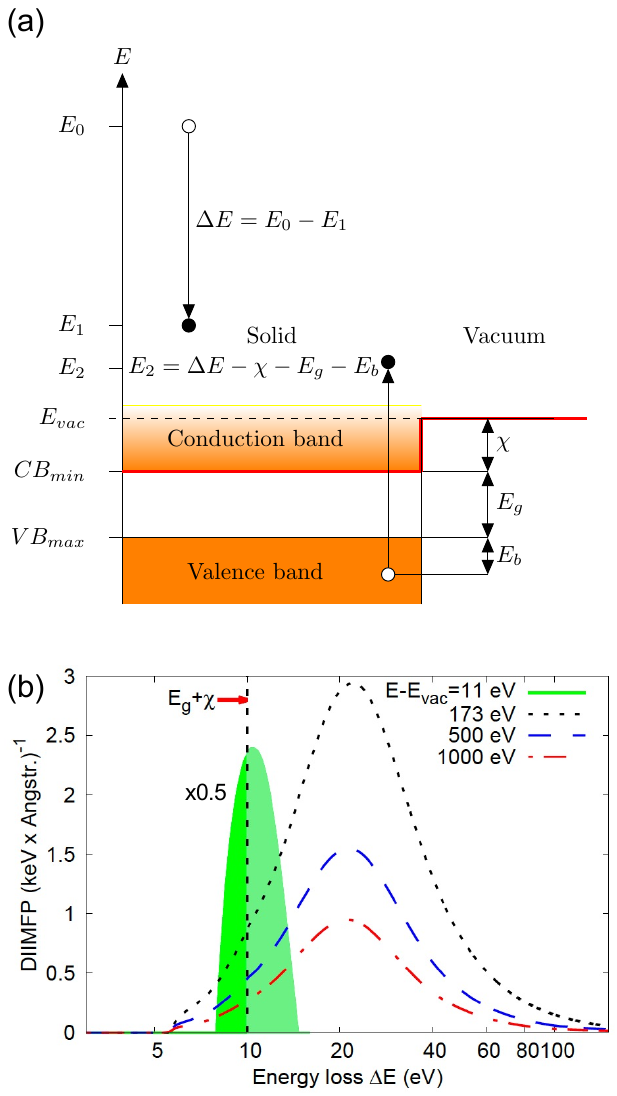}}
\caption{%
(a) Schematic illustration of the simple model for SE-emission from insulating solids ( \protect{\cite{vytbruce}}, see text);
(b)Differential inverse inelastic mean free path (DIIMFP) for PMMA for energies of 11, 173, 500 and 1000~eV  above the vacuum level\protect{\cite{werolga}}. 
}
\label{fdiimfp}
\end{figure}
%

Spectra of  electron pairs correlated in time were measured for electrons with energies of $E_0=$173, 500 and 1000~eV incident on a PMMA surface. 
To avoid charging of the insulator surface, the experimental conditions ensured  that each surface atom on average was hit at the most by one primary electron during acquisition, 
which took about one month for each primary energy (see the SM for experimental details  \cite{SM}).
Fig.~\ref{f500}a shows the  raw data for $E_0=$500~eV on a false colour scale. 
The SE-spectra caused by specific energy losses of the primary are given in Fig.~\ref{f500}b.
Each pixel in the double differential coincidence data in Fig.~\ref{f500}a represents the intensity  
of detected electron pairs: a fast inelastically scattered (primary) electron with energy $E_1$ and a slow (secondary) electron with energy $E_2$ created during the impact of the primary. 
A simple model for the SE-emission process explaining these data is illustrated in Fig.~\ref{fdiimfp}a: in the course of an inelastic scattering process, 
the energy loss $\Delta E=E_0-E_1$ 
of  a primary electron  is transferred to an occupied state in the valence band with  binding energy $E_b$.
The SE-electron liberated inside the solid can be emitted into vacuum if its energy suffices to  overcome 
the surface barrier $U=E_g+\chi$, consisting of the energy gap, $E_g=5.5$~eV \cite{werolga}, and the electron affinity, $\chi=4.5$~eV  \cite{florianprb}.
The yellow curve in Fig.~\ref{f500}a delimits the maximum vacuum energy of a secondary electron created by a given energy loss $E_2=\Delta E -U$.   
The white curve  represents the differential inverse inelastic mean free path (DIIMFP), 
i.e. the distribution  of energy losses in individual inelastic collisions.

 The narrow stripe of high intensity  near the plasmon-resonance just below the yellow curve in Fig.~\ref{f500}a is attributed to a plasmon-assisted  (e,2e)-process  \cite{wercoisec,werhopg,alebruce,weralcoinc}. Multiple plasmon excitation by the primary is responsible for the intensity at  larger 
losses  ($>30$~eV). Here,  the intensity along the $E_2$ axis approximately peaks at  $E_2-E_{vac}=\hbar\omega_p-\chi-E_g\sim 11$~eV (see Fig.~\ref{f500}b), in a process where 
a plasmon decays 
 and  the resonance energy  $\hbar\omega_p$ is transfered to a single   
 solid-state electron near the valence band maximum that overcomes the surface barrier.
 The well-known phenomenon \cite{wercoisec,werhopg,alebruce,weralcoinc} that each energy loss leads to 
 liberation of a single sold-state electron  follows from  the fact that  width of the plasmon 
 feature along the binding energy axis   in the coincidence spectrum  matches the width of the valence band of PMMA (see Ref.~\cite{SM}).  
 The similarity of the coincidence SE-spectra for arbitrary energy loss ranges indicates that the source energy distribution of  SEs depends weakly on the 
energy of the electrons generating them. The reason is that the shape of the  DIIMFP does not significantly  change with the energy of the projectile, which in our case is the  primary electron after  multiple plasmon losses. This  follows from  Fig.~\ref{fdiimfp}b 
\cite{werqsasia} showing  the DIIMFP for various energies calculated from optical data \cite{werolga}.  For projectile energies well above the plasmon resonance of $\hbar\omega_p\sim$21~eV, their 
shape is  practically 
identical.  The maximum energy loss for  11~eV electrons (above vacuum)  is seen to be 15.5~eV, since no allowed states exists at  energies below $E_{vac}-\chi$.
These observations  provide further  evidence for the Markov-type character  of multiple inelastic electron scattering  leading to  SE-emission  \cite{werner2011moves} .

It is perhaps  surprising that the maximum in the SE spectra in Fig.~\ref{f500} is found at 11~eV,  a much higher energy than typically observed in SE-spectra. It should be kept in mind that 
the data in Fig.~\ref{f500} are special in that they constitute SE-spectra emitted as a result of a specific energy loss.
Indeed,  the maximum of the SE-peak in the singles spectra as well as the peak in  the cascade region in the coincidences  is located at $\sim$3.7~eV (not shown, 
  \cite{florianprb}).

\begin{figure}[t]
{\includegraphics[width=0.85\columnwidth,trim={0.2cm 5.0cm 5.cm .cm},clip]{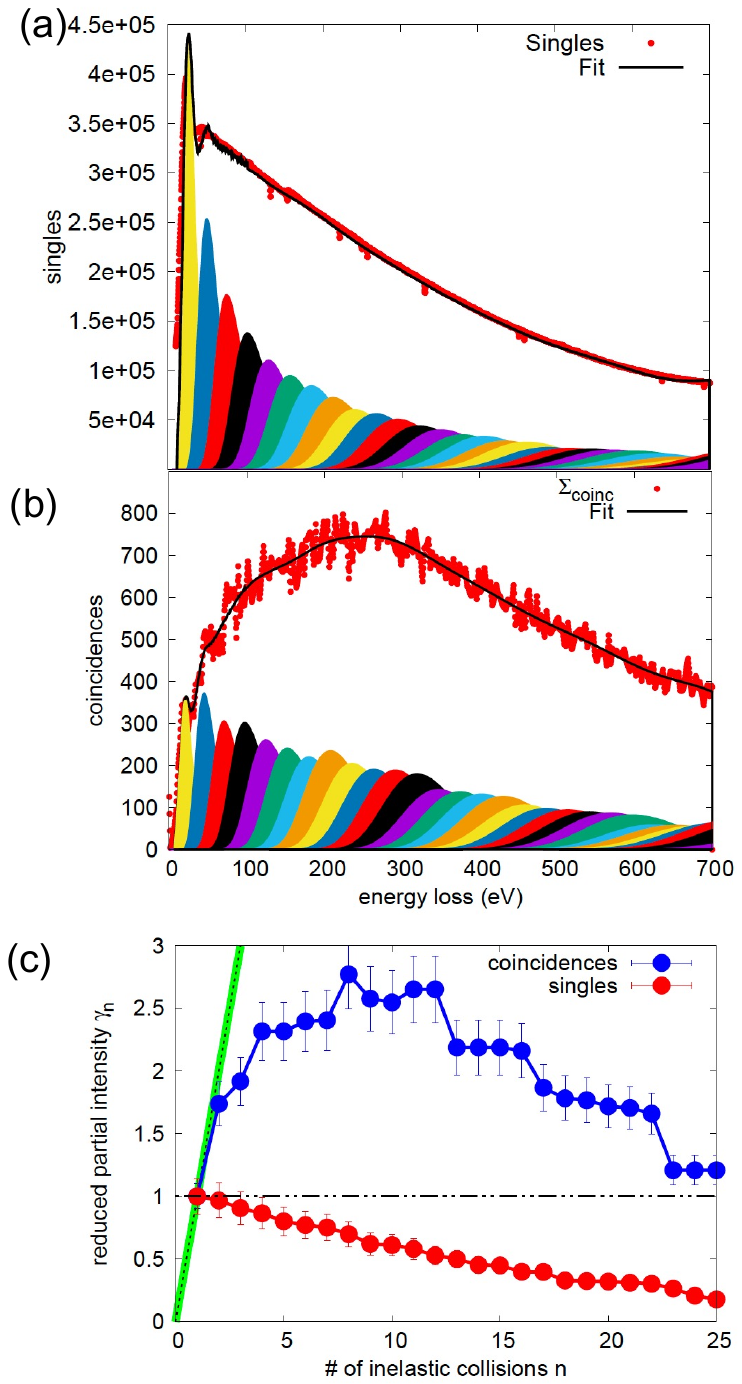}}
\caption{%
(a) Red data points: singles energy loss spectra (acquired during the coincidence run) for a primary energy of $E_0=1000$~eV; (b) corresponding coincidence spectra,
obtained by integrating the double differential data over $E_2$. Black curves are a fit of these data to a linear combination of  
multiple self-convolutions of the DIIMFP, shown by the filled coloured curves; (c) reduced partial intensities $\gamma_n=C_n/C_{n=1}$,
where the quantities $C_n$ are the areas under the filled curves for the spectra shown in (a) and (b). The green line represents the identity, $\gamma_n=n$.}
\label{ffit}
\end{figure}

These observations qualitatively clarify the first stage of the energy dissipation of swift electrons in PMMA: the projectile spends its energy in the course of  multiple plasmon excitations, as illustrated 
by the filled curves in Fig.~\ref{ffit}a and b. Subsequent plasmon decay  induces interband transitions leading to SE emission with energy distributions practically independent of the energy loss of the primary since   the 
shape of the DIIMFP depends very weakly on the projectile energy (as long as it exceeds the plasmon resonance energy).

The intensity of coincidences along  $\Delta E$  in Fig.~\ref{f500}a  is remarkable in that it increases monotonically 
up to an energy of $\sim$150~eV and decreases afterwards.
A similar behaviour was observed for all primary energies and can be seen more clearly for 1000~eV in Fig.~\ref{ffit}a and b: while the intensity in the singles 
energy  loss spectrum (Fig.~\ref{ffit}a), decreases monotonically with the energy loss, the total number of emitted SEs (i.e. the coincidence data integrated over $E_2$, 
Fig.~\ref{ffit}b) exhibits a maximum at  $\Delta E\sim250$~eV. 

The electron energy loss spectrum is a superposition of the $n$-fold self-convolutions of the DIIMFP   \cite{wernerfrontiers,egerton}.
Fitting the spectra to a linear combination of such functions then yields the contribution  of 
$n$-fold inelastically scattered primaries to the spectrum  \cite{werqsasia,wernerfrontiers}. 
The corresponding fits are shown as black curves in Fig.~\ref{ffit}b and c, while the coloured filled 
curves represent the contributions to the spectra of individual $n$-fold energy losses. The areas under these curves correspond, respectively, to the number  of inelastic collisions 
experienced by the  primaries (for the singles spectrum)  and  the number of secondary electrons emitted as a result (for the coincidence spectrum). These 
quantities are referred to as partial intensities, $C_n$  \cite{partintdef}. The reduced partial intensities, $\gamma_n=C_n/C_1$  are presented in Fig.~\ref{ffit}c.

%

For the first few scattering orders, the singles partial intensities are close to unity. It is then expected that the coincidence partial intensities should 
 follow the relationship $\gamma_n^{coi}=n$ (green line in Fig.~\ref{ffit}c) since $n$ energy losses  create $n$ secondary electrons. However, all coincidence partial 
intensities consistently lie below the green line. 
The probability for $n$-fold scattering increases with the travelled pathlength   \cite{werprcsd}, i.e., the average depth at which higher order collisions take place increases 
monotonically with collision number. Then, the decrease of the number of emitted secondary electrons with increasing scattering order, i.e.,  
the deviation of the coincident intensity from the expected behaviour $\gamma_n^{coi}=n$, is attributable to a  corresponding increase  of depth of creation $\langle z_n\rangle$.

\begin{figure}[t]
{\includegraphics[width=0.85\columnwidth,trim={0.2cm 10cm 5.cm .cm},clip]{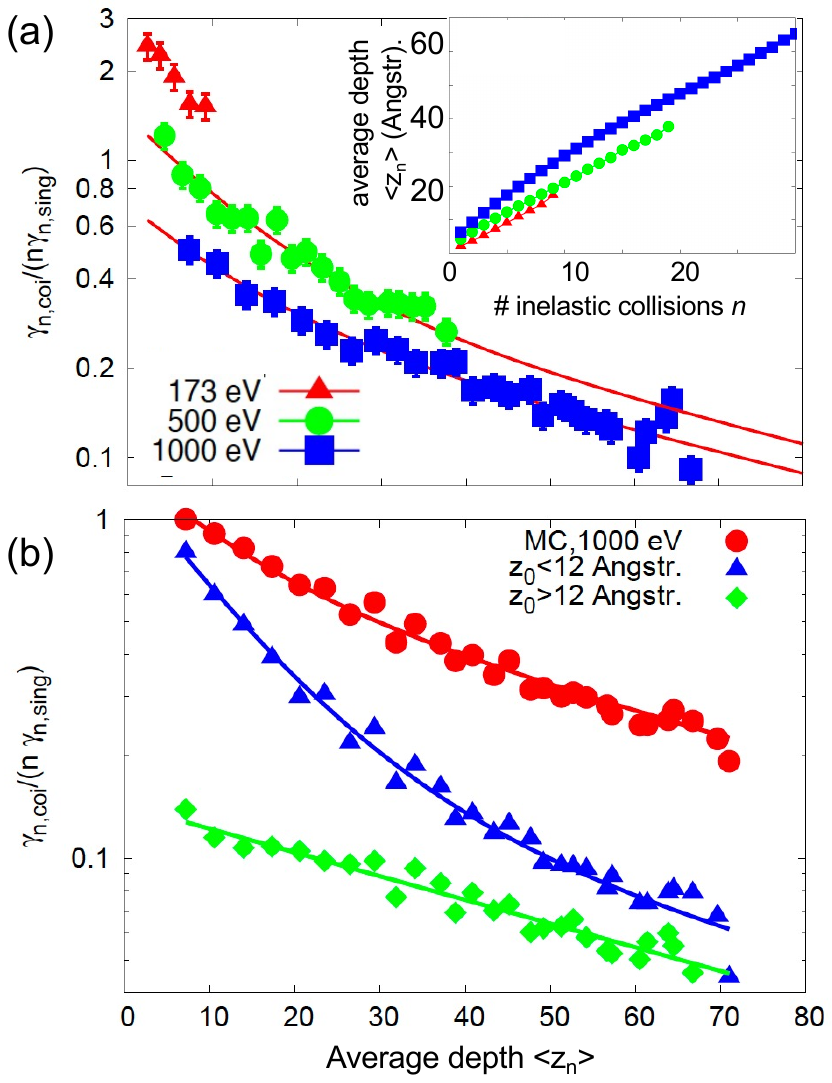}}
\caption{%
(a) Attenuation of the SE yield as a function of depth, $\gamma_{n}^{coi.}/n\times \gamma_{n}^{sing.}(\langle z \rangle)$, for primary energies of 173, 500 and 1000~eV.
The red solid curves in (a) represent a fit to a double exponential 
function (Eq.~\ref{edouble}). The inset shows the MC results for the average depth $\langle z_n \rangle $ at which, on average, $n$ inelastic collisions occur.
(b) Red circles: results of MC model calculations for the quantity shown in (a) (see text). 
The data were  multiplied by a  factor  improving distinguishability of individual curves.}
\label{feal}
\end{figure}

At this stage we invoke the quantitative model for  medium-energy electron-solid interaction  \cite{powjabepes,werjpcrd} to calculate the average depth $\langle z_n\rangle$  at 
which $n$-fold scattering takes place using a Monte Carlo (MC) model (see SM \cite{SM}). 
Since $n$-fold scattering leads to generation of $n$ secondary 
electrons at the corresponding depths, the quantity $\gamma_n^{coi.}/n\times \gamma_{n}^{sing.}$ as a function of $\langle z_n\rangle$ describes the attenuation of SE  
created at a certain depth before they reach the surface. 
This relationship is shown in  Fig.~\ref{feal}a  on a semilogarithmic plot.  The accessible depth ranges are widely 
different for the three considered primary energies but their  depth dependence is satisfactorily described by the same attenuation law, which is clearly not a simple exponential function:
the  solid (red) curves represent a fit to a double exponential function,
\begin{equation}
\alpha_1 \exp(-z/\lambda_1)+\alpha_2 \exp(-z/\lambda_2),
\label{edouble}
\end{equation}  
yielding distinctly different characteristic lengths of  $\lambda_1=(12.0\pm2)\rm{\AA}$ and  $\lambda_2=(61.5\pm11)\rm{\AA}$.

The same analysis was applied to simulated spectra using our MC model.
An essential aspect of this MC model is that deflections in the course of  inelastic collisions are taken into account using a quantum-mechanical approach  \cite{ding89b}.
 The MC-results for 1000~eV are shown by the red  circles in Fig.~\ref{feal}b, along  with a fit (solid red curve) to a double exponential curve with the same characteristic lengths  $\lambda_1$ and $\lambda_2$ 
as in (a).  
The  blue and green points are for subsets of these data 
for depths of origin $z_0$ smaller (triangles, blue)  and larger than 12~$\rm{\AA}$ (diamonds,green).
The solid blue curve is a double exponential function with the same characteristic lengths as above, the solid green curve is a single 
exponential function with characteristic length $\lambda_2$. The MC calculations also yield the mean  vacuum energies in the above two groups  as 
$\langle E_{\lambda_1}\rangle =11$~eV and $\langle E_{\lambda_2}\rangle=4$~eV. 

These results suggest a rather simple chain of processes for the first stages of energy 
dissipation of fast electrons in PMMA. Multiple plasmon excitation 
of the fast primary electron leads to creation of secondaries with energy distribution peaking around $E-E_{vac}=\hbar\omega_p-E_g-\chi\sim11$~eV. 
During the transport to the surface such an electron has a significant  probability 
to undergo an inelastic collision:   the area under the curves  for 11 and 1000~eV in Fig.~\ref{fdiimfp}b is of the same order of magnitude. 
Let the corresponding characteristic 
length be denoted by   $\lambda_1$.  If such a  first generation "11eV"-secondary electron is created at a depth  larger than $\lambda_1$  it is likely to suffer another energy loss before escape. In case 
this energy loss  is smaller than the surface barrier ($\Delta E < U$), it is transferred to an electron in the valence band, the latter  (liberated) electron can only be promoted to a 
hot-electron state in the conduction band below the vacuum level. It cannot escape into vacuum. The inelastically scattered electron 
itself will have an energy just above vacuum after the collision. 

The other case when the energy loss of the first generation secondary electron exceeds the surface barrier ($\Delta E>U$), leads to a situation where in the final state, the role of 
the scattered and liberated electron is reversed: the scattered electron  will be a hot electron below the vacuum level, while the liberated electron will have a positive vacuum 
energy and can be emitted as a secondary electron (of the second generation). In both cases, the energy of the  electrons with a positive vacuum energy will be small (typically of 
the order of a few eV above vacuum) and their IMFP  will be large due to the limited availability of final states in further scattering processes. Hence the characteristic length for 
attenuation for the second generation  ($\lambda_2$) will also be large (green diamonds, $z_0>12~\rm{\AA}$). 

If a first generation "11eV"-electron is created at a depth smaller than $\lambda_1$ (blue triangles, $z_0<12~\rm{\AA}$) it can escape without further loss if its initial direction points outward, otherwise it will scatter 
and  belong to the $\lambda_{2}$-group thereafter.
The mechanism outlined above  corresponds exactly to the results shown in Fig.~\ref{feal}b.

\begin{figure}[t]
{\includegraphics[width=1.\columnwidth,trim={1.2cm 6.5cm 13.5cm 01.cm},clip]{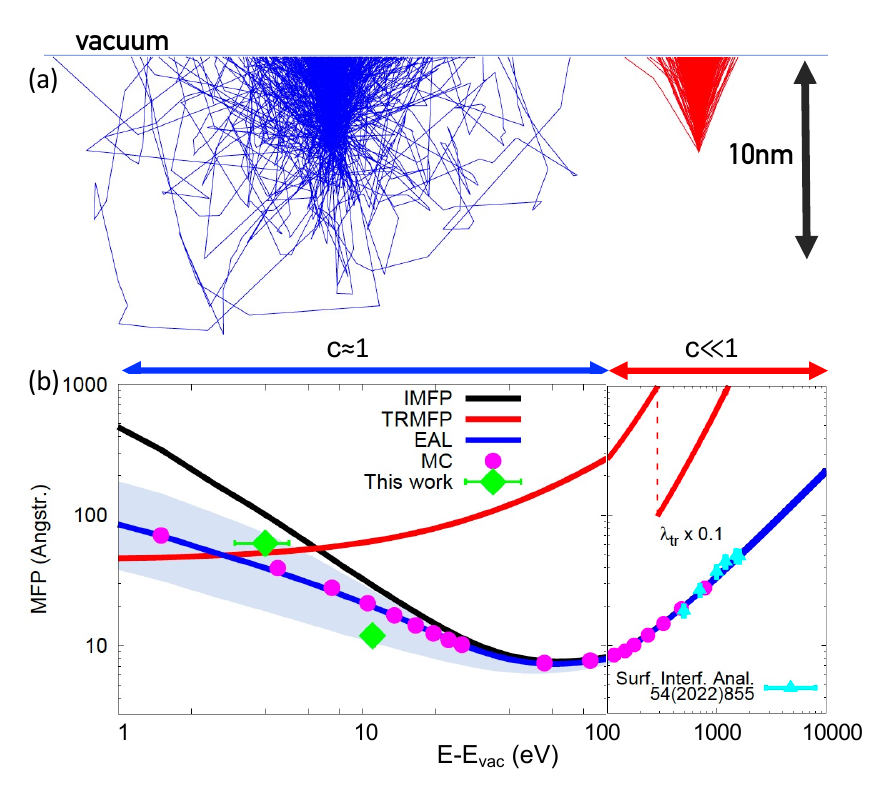}}
\caption{%
(a) Trajectories of electrons emitted isotropically at a depth of 5nm and reaching the surface without energy loss. 
 Left, (blue):  single scattering albedo $c\approx1$; Right (red): $c\ll1$.
(b) Electron inelastic mean free path (IMFP, black,  \protect{\cite{shinotsuka2022}}), transport mean free path (TrMFP, red) and effective attenuation length (EAL, blue, Eq.\ref{eeal}). The (magenta) circles are results of MC model calculations for the EAL, the (cyan) triangles are earlier experimental data  \protect{\cite{werimfp2022}}. The  green diamonds represent the results for $\lambda_{1,2}$ derived from the data in Fig.~\ref{feal}a. The blue shaded  region represents the  uncertainty in the EAL when the TrMFP is increased/decreased by a factor of three. 
}
\label{fbarb}
\end{figure}
%

In the framework of linear transport theory, the expression for the effective attenuation length (EAL), $\lambda_a$, that takes into account the combined influence of energy fluctuations (inelastic scattering) and momentum relaxation (deflections) is given by \cite{case,chandrasekhar,davison}: 
\begin{equation}
\lambda_a=\frac{\lambda_i \lambda_{tr}}{\lambda_i +\lambda_{tr}}\nu_0=\lambda_{tr}c\nu_0
\label{eeal}
\end{equation}
where the single scattering albedo is given by $c=\lambda_i/(\lambda_i+\lambda_{tr})$ and the quantity $\nu_0$ is the positive root  of the characteristic equation
\begin{equation}
\frac{2}{c}=\nu_0\ln
  \frac{\nu_0+1}{\nu_0-1} 
\label{enu0}
\end{equation}

In the medium energy range, the TrMFP exceeds the IMFP by a significant 
factor, yielding a  value for $\nu_0$ very close to unity and the EAL is slightly smaller than the IMFP, the difference being  $\sim$10\% or less. For small values of $c \ll 1$, 
the particle will approximately move along a straight line  and the attenuation is dominated by the IMFP (see Fig.~\ref{fbarb}a).
For low energies,  as the TrMFP assumes values of the order of the IMFP or less and the influence of momentum relaxation becomes more pronounced, the EAL and IMFP 
are essentially different. For values of $c\sim1$, many deflections occur before an inelastic process can take place. 

Identifying   $\langle E_{\lambda_{1,2}}\rangle$ as the  energies associated with the characteristic lengths  of the two stages of the energy dissipation process, $\lambda_{1,2}$  are shown as 
green diamonds in Fig.~\ref{fbarb}b and are compared with the mean free path for inelastic scattering $\lambda_i$ (IMFP \cite{shinotsuka2022}) and momentum relaxation  $
\lambda_{tr}$ (TrMFP \cite{salelsepa}) as well as the effective attenuation length $\lambda_a$ 
according to Eqn.~\ref{eeal}. 

The (magenta) circles in Fig.~\ref{fbarb} were calculated with the MC-technique and agree quantitatively with 
 Eq.~\ref{eeal}, which  adequately accounts for the relative importance of energy fluctuations and momentum relaxation. 
The present results for $\lambda_1$ and 
$\lambda_2$ differ by more than a factor  of two with the  IMFP and agree significantly better with Eq.~\ref{eeal},
underscoring  the importance  to adequately account for the combined influence  of collective excitations and momentum relaxation.

In summary,  the energy dissipation process of fast electrons in PMMA begins with multiple plasmon excitation of the primary.
Plasmon decay induces interband transitions acting as sources of SEs. The SE source energy distribution depends weakly on
the energy of the primary since the shape of the DIIMFP is very similar for any projectile energy (above the plasmon resonance).
The subsequent analysis  identifies two groups in the SE cascade which correspond to different stages in the energy dissipation process.
The  associated characteristic lengths  for electron beam attenuation have been determined using 
the quantitative model for medium energy transport to calculate the corresponding depth scale.
Comparison of the characteristics length $\lambda_{1,2}$ with the universal curve, Eqn.\ref{eeal},
suggests that the transport of low energy electrons can be described by the same physical laws as those in light scattering in interplanetary nebulae,
impressively demonstrating the scaling of physical laws over 26 orders of magnitude.
There is  consensus in the community working on  attenuation of electron beams at medium energies, 
that adopting elements of linear transport theory to compare experiment and theory constituted an essential step \cite{powjabepes,wernerfrontiers}.
While for medium energies, the influence of momentum relaxation leads to a rather minor correction of the order of $\sim$10\%,  it plays an essential role in the electron 
transport for low energies. The reasonable agreement in Fig.~\ref{fbarb}b between the experimental attenuation lengths $\lambda_{1,2}$ and Eqn.\ref{eeal}   indeed 
suggests that the scientific debate on low energy electron attenuation should explore  the merits  of linear transport theory at the earliest stage possible.

{\bf Acknowledgments}\\
The computational results  have been achieved using the Vienna Scientific Cluster (VSC).
TU Wien Bibliothek is acknowledged for financial support through its Open Access Funding Programme.

\bibliography{./bibtex/Allrefnewest,./bibtex/LEE,./bibtex/DNA,./bibtex/EUV,./bibtex/supp}

\newpage 
.
\newpage 
.

%
\begin{center}
{\Large Supplemental Material for {\bf "Energy dissipation of fast electrons in polymethylmetacrylate (PMMA): Towards a universal curve for electron beam attenuation in solids for energies between 0\,eV and 100\,keV"}}\\
{\it Wolfgang S.M. Werner, }%
{\it Florian Simperl, }%
{\it Felix Bl{\"o}dorn, }%
{\it Julian Brunner, }%
{\it Johannes Kero, }%
\newline
{Institut f\"ur Angewandte Physik, Technische Universit\"at  Wien, Wiedner Hauptstra{\ss}e 8-10/E134, A-1040 Vienna, Austria}
\newline {\it Alessandra Bellissimo, }
\newline  {Institut f\"ur Photonik, Technische Universit\"at  Wien, Gu{\ss}hausstra{\ss}e 27-29/E387, A-1040 Vienna, Austria}
\newline {\it Olga Ridzel}
\newline {Theiss Research, 7411 Eads Ave., La Jolla, CA 92037-5037, USA.}
\end{center}

\section{\label{sec:Experiment}Experimental}

\begin{figure}[t]
{\includegraphics[width=1.\columnwidth,trim={0.1cm .10cm .1cm .1cm},clip]{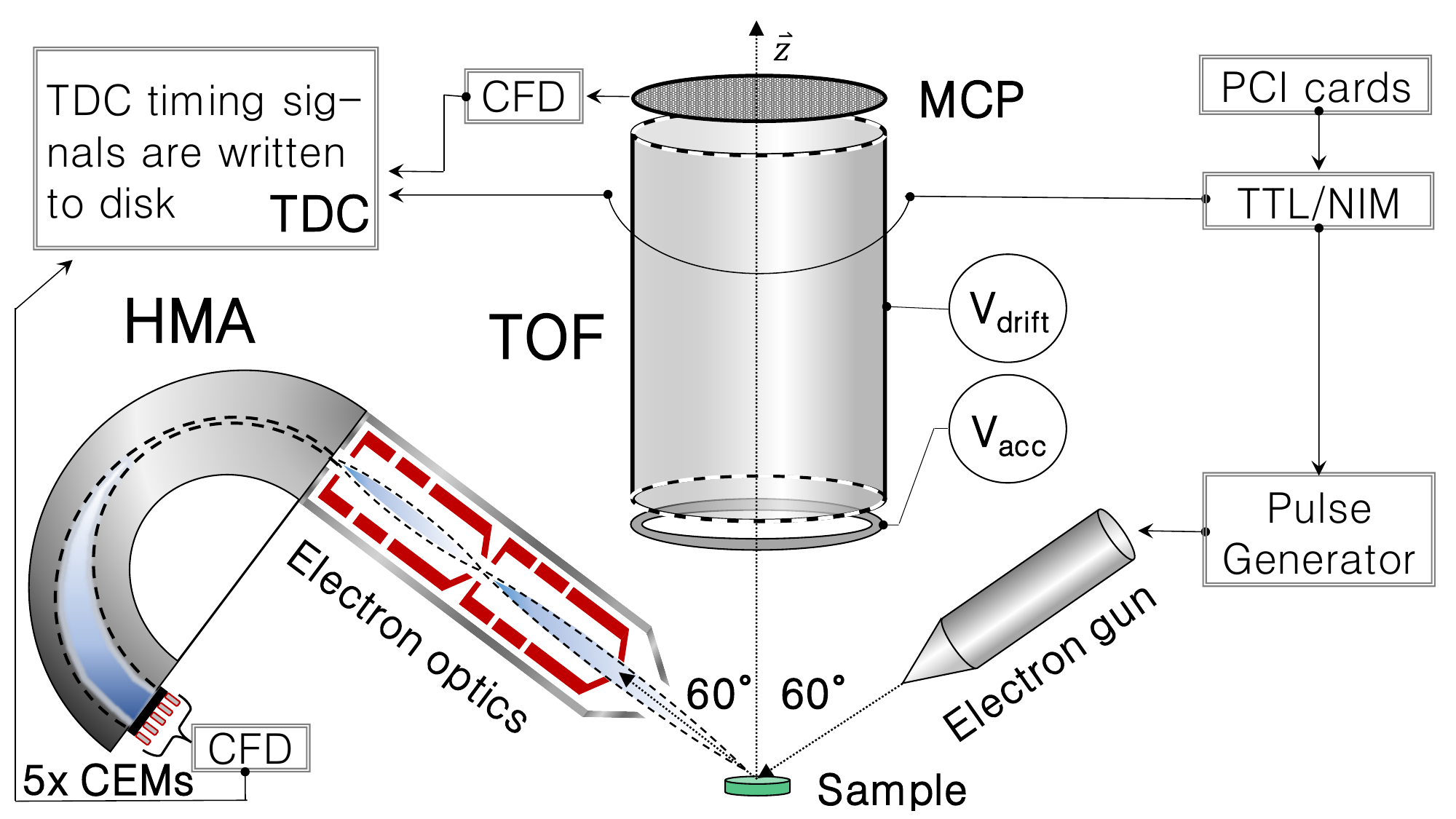}}
    \caption{Schematic view of the SE2ELCS (secondary electron-electron energy loss coincidence) spectrometer \protect{\cite{SE2ELCS2020}}}
    \label{fig:schem_exp}
\end{figure}

  A schematic illustration of the experimental setup is shown in Fig.~\ref{fig:schem_exp}: 
An electron gun provides a stable low-current electron beam  incident on the surface and electron 
pairs leaving the surface as a result are detected by a hemispherical analyser 
(HMA)  and a time-of-flight analyser (TOF). 
The experiment is conducted in a UHV-chamber with a pressure during the experiment not exceeding 2$\times 10^{-10}$~mbar.
The arrival times in either analyser  are written to 
disk and coincidences are retrieved from the data  retrospectively. The HMA
 is equipped with 5 channeltron detectors and is operated in the constant analyser energy  mode with $E_{pass}=200$\,eV (energy resolution 5~eV) for coincidence measurements and $E_{pass}=20$\,eV 
(energy resolution 0.5~eV) for singles spectra. The electrons 
in the TOF analyser are detected by a stack of two multi-channel plates  and a delay-line anode.  The energy resolution of the TOF  is a tenth of an eV at an energy of 1eV and tens of an eV at energies of a few hundred 
 eV. The entrance aperture of the TOF is kept at a potential of +5~V to accelerate slow electrons towards it. The angle of incidence and detection with the HMA decsribe  an angle of 60$^\circ$ with the surface normal, the TOF-axis is parallel to the surface normal.

 During the coincidence measurements, the Kimball Physics ELG-2 electron gun is operated with a continuous  beam  at a low 
 current of $\sim$1\,pA. Before a coincidence measurement, a pulsed electron beam is used to calibrate the  time it takes  an electron with a given energy to leave  the sample  
  and to reach  one of the channeltrons in the HMA. This is repeated for each energy used later  during the coincidence run.
 Within the experimental resolution, the two electrons in the pair are emitted simultaneously since the duration of the emission process (of the order of femtoseconds) is orders of 
 magnitude smaller than the net time resolution of our experiment, which amounts to a few nanoseconds. For the coincidence measurement, we then use the calibrated  flight times of  electrons in the HMA to trace back the starting time of the pair, 
 yielding the TOF-flight time in spite of the use of a continuous electron beam.
 The above procedure is advantageous compared to pulsed beam experiments in that it allows one to use larger currents and gives rise to higher coincidence rates.  
 With this setup, a histogram of arrival time differences (between electrons arriving in the HMA and TOF) exhibits a peak of true coincidences superimposed on a flat background of random coincidences, which is subtracted.   When the current is increased, the background intensity increases quadratically, while the intensity in the  peak increases linearly, proving that it is made up of true coincidences \cite{jensen}. 
  \newline

Coincidence measurements were conducted for  three different primary energies $E_0 - E_{vac}$, i.~e.~173\,eV, 500\,eV and 1000\,eV. 
We use a 1x1\,cm Po\-ly\-me\-thyl\-me\-tha\-cry\-late (\text{C$_5$H$_8$O$_2$}) sample with a thickness of 50\,nm on a silicon substrate. The position of irradiation is changed 
every 24 hours by 1\,mm to reduce surface charging. 
Under these conditions, each atom on the surface on average is hit by one primary electron or less during acquisition.
Total acquisition time for each measurement amounted to about 1 month. 
The number of detected electrons in the HMA during the coincidence run is recorded as "singles"-spectrum and used in the analysis in Fig.~4 of the main text to determine the 
fraction of secondary electrons generated at a certain depth  which are eventually emitted from the surface. In this way, spurious influences due to e.g., drift in the beam current, the transmission function of the analyser, etc., are eliminated.
%

\section{Data Interpretation}
We designate the scattered and ejected electron by the indices "s"  and "e" , while in the main text we merely distinguish between events where electrons are detected in 
analyser 1 and 2  and label the energy scales accordingly. 
This is strictly speaking correct and necessary because of the indistinguishability of electrons but generally identifying electron 1 with the scattered (fast) electron and electron 2 with the ejected (slow) electron will be correct in most cases.

The binding energy $E_b$ of the bound electron before it is liberated in the collision with the primary is found by requiring that the energy loss of the primary electron  $\Delta E=E_0-E_s$ --where the index "0" indicates the primary electron-- is used to liberate the bound electron from the solid, by overcoming the surface potential barrier $U$, and that it is ultimately ejected from the solid with an energy $E_e$ above the vacuum level:
\begin{equation}
\label{econs}
\Delta E=E_0-E_s=E_e+U-E_b
,
\end{equation}
where the  binding energy is counted from the top of the valence band and is negative. The surface potential barrier $U$ is the sum of the band gap energy $E_g$
and the electron affinity $\chi$.
The binding energy of the solid state  electron before liberation  follows from the above as:
\begin{equation}
\label{econs}
E_b=-E_0+(E_e+E_s)+U,
\end{equation}
or, in other words, the spectrum along the energy sum axis in fact represents the binding energy spectrum of the solid state electrons.  

 The double-differential (e,2e)-coincidence spectrum presented in Fig.~1(a) in the main text displays the intensity of time-correlated electron pairs given as a function of 
the energy loss of the fast electron  and the energy of an emitted electron. The energy loss directly results from the energy transfer of a primary electron that undergoes an 
inelastic collision (with {$\Delta E=E_0 - E_s$)  in the sample, releasing it as a secondary electron with kinetic energy $E_e=\Delta E- U+E_b$, either via a direct knock-on collision 
with a single solid state electron or after excitation and decay of a collective excitation, such as a plasmon. Hence, each pixel shown in the spectrum of Fig.~(1)a. represents the intensity of one 
such correlated electron pair. 

\begin{figure}[t]
{\includegraphics[width=1.\columnwidth,trim={0.2cm 11.0cm 15.cm .cm},clip]{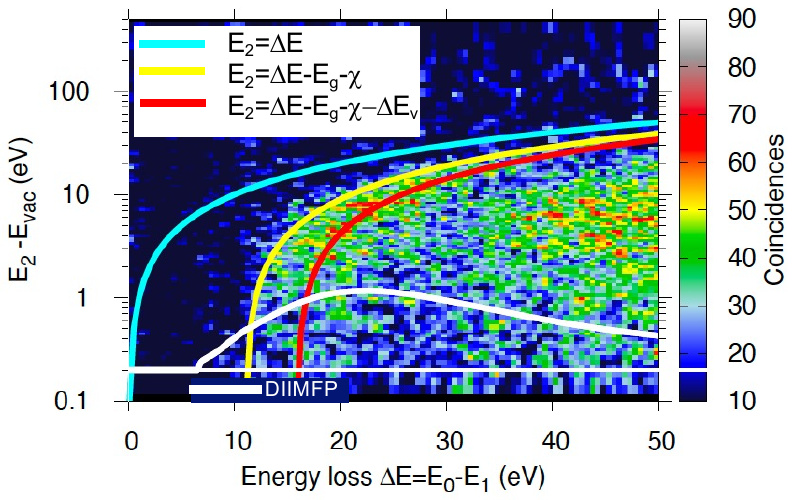}}
    \caption{Portion of Fig.~1a in the main text around the plasmon loss feature. The red curve indicates the bottom of the valence band according to the value of $\Delta E_v$ in Ref.~\protect{\cite{shinotsuka2022}}.}
\label{f500sm}
\end{figure}
%

The fact that each energy loss results in the liberation of exactly one solid-state electron is one of the central ideas of the present work, 
since it allows to construct the depth dependent attenuation curves shown in Fig.~4a in the main text. Only in this case and only without depth dependent attenuation of the 
outgoing beam, the reduced partial intensities in the coincidence loss spectrum should follow the green curve,  $\gamma_n^{coi.}=n$, in Fig.~3c (in the main text). 
This issue  has been discussed previously in several  works \cite{wercoisec,werhopg,alebruce,weralcoinc} and is clarified here for completeness in Fig.~\ref{f500sm},
which shows the details around  the plasmon feature in Fig.~1a (in the main text).  The yellow curve 
is defined by $E_e=\Delta E-E_g-\chi$ and represents the  top of the valence band.	 
The red curve indicates the binding energy corresponding to the bottom of the valence band  
according to the value of $\Delta E_v=15.8$~eV found in Ref.~\cite{shinotsuka2022}.
 If more than one electron would be released for a given energy loss $\Delta E$ (e.g. in the plasmon feature), then, on any realistic  model for the energy sharing between the 
 released electrons, one would expect intensity at all energies $E_e<\Delta E-U$, for the considered energy loss $\Delta E$.
  This is clearly not the case. Rather, the range of binding 
 energies in the plasmon feature matches the width of the valence band quite well. 
 
 The above applies to the single scattering region ($\Delta E=10\sim 30$~eV). For larger energy losses consecutive (multiple) plasmon features overlap leading to intensity at any energy $E_e$ 
 
 Coherent multiple plasmon excitation and decay, in which $k$ plasmons are excited and the total energy loss $\Delta E_{tot}=k\hbar \omega_p$ is transferred to a single ejected 
 electron  in a  coherent process would give rise to intensity near $(E_s,E_e)=(E_0-k\hbar \omega_p,E_0-k\hbar \omega_p-U)$. This can neither be observed in Fig.~1a (of the 
 main text) nor has it been observed in previous works \cite{wercoisec,werhopg,alebruce,weralcoinc} which include measurements on single crystals \cite{alediss}. Rather, at 
 energy losses corresponding to multiple plasmon excitation, a peak is observed with its maximum always at $E_e=\hbar\omega_p-U\sim 11$~eV (see Fig.1b in the main text). 
 Then, multiple plasmon excitation is governed by a Markov-type process and each energy loss leads to liberation of a single solid-state electron.

\section{\label{sec:ScatteringTheor}Physical model for electron scattering} 
The findings presented in the main text and discussed in the following section exclusively pertain to polycrystalline or amorphous solids. In this case, the off-diagonal 
elements of the density matrix, responsible for quantum mechanical interference effects, can be neglected.
This leads to a Boltzmann-type kinetic equation describing electron transport which was implicitly used throughout the present work \cite{schattwertem,werschatteels}. However, 
a quantum mechanical approach was used to calculate all interaction parameters as described below. \newline

Assuming that spin can be neglected, the electron transport in solids depends on processes that alter the direction and energy of electrons propagating in solids, described in 
terms of elastic and inelastic  scattering. Elastic scattering is defined by the interaction of an electron with the screened Coulomb potential of the nucleus leading to a deflection  
and a small recoil energy loss which is negligible for the present work by virue of the large  mass difference between the scattering partners. 
Inelastic scattering describes the process of electron interaction with solid state electrons leading to appreciable energy loss and a small (but in the present case non-negligible) momentum transfer. 
The differential inverse  inelastic mean free path (DIIMFP) $W_{in}(\omega,E)$ describes the probability for an 
electron with energy $E$ to lose the energy $\hbar\omega$ in a single inelastic collision. Based on the model for non-conductors by 
Tosatti and Parravicini \cite{diimfp_tos_parra}, several authors \cite{boutboul1996,shinotsuka_imfp} express the  DIIMFP for semiconductors and insulators
in terms of the dielectric function $\epsilon(\omega,q)$ as
\begin{equation} \label{Eq:DIIMFP}
    W_{in}(\omega,E) = \frac{1}{\pi(E - E_g)} \int_{q-}^{q+} \Im \left[-\frac{1}{\epsilon(\omega,q)}\right]\frac{dq}{q},
\end{equation} 
where  $E$ is the incident energy and   
the boundaries for the momentum transfer  $q$ are given as $q_{\pm} = \sqrt{2(E-E_g)} \pm \sqrt{2(E-E_g-\omega)}$.
Here and below, atomic units are   used ($\hbar=m_e=e\equiv 1$).
 The IMFP is obtained by integrating the DIIMFP over all allowed energy losses  \cite{shinotsuka_imfp}:
\begin{equation}
    \lambda_{in}(E)^{-1}= \frac{1}{\pi (E-E_g)}\int\limits_{E_g}^{E-(\Delta E_v+E_g)} {d\omega} \int\limits_{q_-}^{q_+} \Im \left[\frac{-1}{\epsilon\left(\omega,q\right)} \right] \frac{dq}{q},
    \label{eimfp}
\end{equation}
The lower integration boundary in Eq.~\ref{eimfp} is defined by the smallest possible loss which corresponds 
to a HOMO-LUMO transition, i.e. $\omega_{min}=E_g$. The upper integration boundary is defined by the lowest available state for the primary electron after the collision, i.e., the 
bottom of the conduction band: $\omega_{max}=E_-(\Delta E_v+E_g)$.

  For low energy electrons, deflections due to inelastic collisions become important and we find that the classical approach leads to unphysical spectral shapes. Hence, we rely on the formula given by Ding and Shimizu \cite{ding89b} for the distribution of scattering angles associated with a given energy loss $\omega$
\begin{equation}
\frac{d^2\lambda_{in}^{-1}}{d\Omega d\omega} =  \frac{1}{\pi^2 q^2} \sqrt{1-\frac{\omega}{E - E_g}}  \Im \left[\frac{-1}{\epsilon\left(\omega,q\right)}\right],
\label{imfp_angle2}
\end{equation}
where
\begin{equation}
\label{qtoangle2}
\frac{q^2}{2} = 2(E-E_g - \omega) - 2\sqrt{(E-E_g)(E - E_g - \omega)}\cos \theta
\end{equation}
and $\theta$ is the polar scattering angle. To evaluate the above quantities, we used the dielectric function given  
by Ridzel et al. \cite{ridzel_pmma_die}  employing a quadratic dispersion.

Deflections occuring  during elastic scattering were modelled using the values for the differential Mott cross section for elastic scattering 
$d\sigma(\theta)/d\Omega$
provided by the ELSEPA package \cite{salelsepa}. The elastic mean free path $\lambda_e$ is obtained by integrating the cross section over the unit sphere. 
The transport mean free path $\lambda_{tr}$ in essence gives the characteristic length for momentum transfer \cite{werqsasia}:
\begin{equation}
\frac{1}{\lambda_{tr}}=N_a \int\limits_{4\pi} \frac{d\sigma(\theta)}{d\Omega}(1-\cos\theta)\sin\theta d\theta d\phi,
\end{equation}
where $N_a$ is the density of scattering centers.

For medium energies, above a few hundred eV, the inelastic interaction characteristics have been extensively tested, mainly by comparison of Monte Carlo model calculations 
and experiments \cite{powjabepes,wernerfrontiers}. The uncertainty in IMFP values for medium energies is nowadays believed to be better than 10\%, while for lower energies, 
the uncertainty is essentially unknown. The same can be said to be true for the elastic interaction characteristics, such as the transport mean free path. Here, for lower energies, the commonly made assumptions that exchange and polarisation effects are negligible makes the resulting quantities less reliable.

Finally, an electron reaches vacuum only if it can overcome the surface potential barrier represented by a step potential in the one dimensional Schr\"odinger equation. Two cases need to be distinguished  for the 
transmissions function $T(E,\phi)$: either the electron  crosses the surface and is refracted, or is internally reflected back:
\begin{equation}
    T(E,\phi) = 
    \begin{cases}
       \displaystyle \frac{4 \sqrt{1-\frac{E_g+\chi}{E \cos^2\phi}}}{\left(1+\sqrt{1-\frac{E_g+\chi}{E \cos^2\phi}}\right)^2}, & \text{if}\ E\cos^2\phi > E_g+\chi \\
        0, & \text{if}\ E\cos^2\phi \leq E_g+\chi  
    \end{cases}
    \label{eq:transmission}
\end{equation}
where $\phi$ is the angle of the electron relative to the  surface normal and the barrier height  is taken to be the electron affinity $\chi$, 
as  illustrated in Fig.~2(a) in the main text.

\section{\label{sec:MCSim}Monte Carlo Simulation}
In general, MC simulations allow one to approximate the solution to the complicated multidimensional transport equations by statistical sampling. The present algorithm essentially follows algorithms which can be found in the literature (e.g. \cite{werqsasia,ding89b}). The electron trajectory is assumed to consist of piece-wise straight line segments in between scattering processes. The step lengths are sampled analytically using the inverse cumulative distribution method applied to an exponential distribution:
\begin{equation}
    s=-\lambda_{tot}\ln(\xi)\\ 
\end{equation}
where  $\lambda_{tot}={1}/({\lambda_{in}^{-1}+\lambda_{el}^{-1}})$,  $\lambda_i$ and $\lambda_e$ are the inelastic and elastic mean free path, respectively, and $\xi$ is a uniform random number on the interval $(0,1]$.
After travelling a step, the position is updated and  another random number  $\xi$  is used to decide whether the scattering process is elastic ($\xi < \lambda_{tot}/\lambda_{el}$) 
or inelastic ($\xi > \lambda_{tot}/\lambda_{el} = 1-\lambda_{tot}/\lambda_{in}$). 
To generate  stochastic values for the energy loss, scattering angle, etc., the corresponding distributions, discussed in the previous section, are sampled 
using the accept and reject method.
The slowing-down process during the electron transport, leads to a  change of scattering characteristics with the projectile's energy. After each inelastic process the relevant 
parameters are updated accordingly using extensive lookup tables.

 Each inelastic scattering process leads to an electronic transition from an occupied state in the valence band with binding energy $E_b$ to an available 
unoccupied state in the conduction band, as described in the main text. We assume that after each inelastic collision, the energy loss and change in momentum of the primary  
are transferred to a single secondary electron  in the valence band (assuming a width of the valence band $\Delta E_v=15.8$~eV\cite{shinotsuka2022}), either in a direct knock-on collision with a solid state electron, or after decay of a collective excitation, e.g., a plasmon. 
 At the solid-vacuum interface, the electron energy 
($E_{sol} = E_{vac} + \chi$) and direction ($\sin\theta_{vac} E_{vac} = \sin\theta_{sol} E_{sol}$) are updated as explained in the previous section. If the maximum angle defining 
the so-called "escape cone" is exceeded, the electron suffers a total internal reflection instead of escaping to vacuum.

The excited secondary electron can reach its final state in vacuum if the energy loss of the primary electron is large enough to overcome the surface barrier.
Whenever a SE is generated, all the information pertinent to this electron is stored on a stack until its trajectory is terminated, 
i.~e.~until the electron is either detected or abandoned (surpassing the maximum simulation depth, reaching minimum escape energy in solid or missing the detector in vacuum).
The SE cascade is simulated by keeping track of the additional energy losses of the previous-generation secondary electrons. The algorithm terminates if the desired number of trajectories have been simulated.

The above algorithm was used in the present work for three main goals: (1.) to calculate the  average depth $\langle z_n \rangle$ at which $n$-fold inelastic scattering (i.e. 
creation of $n$ secondary electrons) takes place (inset of Fig.~4a in the main text); (2.) simulation of the singles and coincidence spectra  (results displayed in Fig.~4b in the 
main text); and (3.) calculation of the energy corresponding to the groups of electrons with attenuation length $\lambda_{1,2}$ (green diamonds in Fig.~5b in the main text).
We trust the results for $\langle z_n \rangle$ to be realistic within 10\% since these model calculations only concern the transport of the primary electron, which is believed to be 
quantitatively understood.

To simulate the coincidence data, for each escaping electron that reaches the detector, its history is stored in the  detector stack. After finishing a given trajectory,  
all $N$ detected electrons are collected into all possible pairs ($N(N-1)/2$ combinations), which are subsequently added as a contribution to a two-dimensional 
histogram, representing the double-differential coincidence spectrum. Since this involves both penetration of the primary into the surface as well as generation and emission of 
secondaries, for which the transport parameters are far less reliable, we cannot make a realistic error estimate. It should be nonetheless noted that an attempt to simulate the 
data in Fig.~4 in the main text using a classical description of deflection angles in inelastic collisions gave results which were qualitatively different from experiment.

Concerning the calculation of the energies $\langle E_{\lambda_1,\lambda_2} \rangle$, one can say that the error bars shown in Fig.~5 of the main text (which are difficult to 
discern since they are about the same as the size of the green diamonds),  are believed to give a realistic estimate of the uncertainty. They were calculated from the width of the 
resulting distribution of energies  $E_{\lambda_1,\lambda_2}$ and amounted to about $\Delta(E_{\lambda_1,\lambda_2}-E_{vac})\sim 1$~eV in both cases.

\end{document}